# Lessons learned from (and since) the Voyager 2 flybys of Uranus and Neptune

Heidi B. Hammel*

*Association of Universities for Research in Astronomy, Inc. (AURA), 1331 Pennsylvania Avenue NW, Suite 1475, Washington, DC 20004, USA, ORCID: 0000-0001-8751-3463*

# Summary

More than 30 years have passed since the Voyager 2 fly-bys of Uranus and Neptune. This paper outlines a range of lessons learned from Voyager, broadly grouped into "process, planning, and people." In terms of process, we must be open to new concepts, whether new instrument technologies, new propulsion systems, or operational modes. Examples from recent decades that could open new vistas in exploration of the deep outer Solar System include the Cassini Resource Exchange and the "sleep" mode from the New Horizons mission. Planning is crucial: mission gaps that last over three decades leave much scope for evolution both in mission development and in the targets themselves. The science is covered in other papers in this issue, but this paper addresses the structure of the US Planetary Decadal Surveys, with a specific urging to move from a "destination-based" organization to a structure based on fundamental science. Coordination of distinct and divergent international planning timelines brings both challenges and opportunity. Complexity in the funding and political processes is amplified when multiple structures must be navigated; but the science is enriched by the diversity of international perspectives, as were represented at the Ice Giant discussion meeting that motivated this review. Finally, the paper turns to people: with generational-length gaps between missions, continuity in knowledge and skills requires careful attention to people. Lessons for the next generation of voyagers include: how to lead and inspire; how to develop perspective to see their missions through decades-long development phases; and cultivation of strategic thinking, altruism, and above all, patience.

*Author for correspondence (hbh@alum.mit.edu).

# Setting the Stage

In this opinion piece, I review the lessons learned from the past several decades of exploration in the fields of astrophysics and planetary science that are of relevance to fulfilling the goals of an Ice Giant mission. Before delving into specific lessons below, I share two overarching lessons which I will return to throughout this paper. First is a lesson from the past that bears repeating today. This "Lesson 0" comes from none other than Michelangelo (**Fig. 1**). One of his more profound and cautionary sayings has direct relevance to those of us who seek to explore the most distant and inaccessible objects: "The greatest danger for most of us is not that our aim is too high and we miss it, but that it is too low and we reach it." I urge my colleagues to aim high, rather than low. At the scale in which most of us aspire to work, revolutionary ideas and missions drive budgets, not the other way around.

The other overarching lesson I call Lesson 1 (**Fig. 2**): "When your mission costs above $1B, science is necessary but not sufficient." There is nothing definitive about "$1B" but experience has consistently shown that once you reach this level of visibility within the U.S. civilian science environment, other considerations become relevant (these may include politics, industry support, personalities, technology needs, agency priorities, NASA Center priorities, etc.). The higher the cost of a mission, the more important those other considerations become.

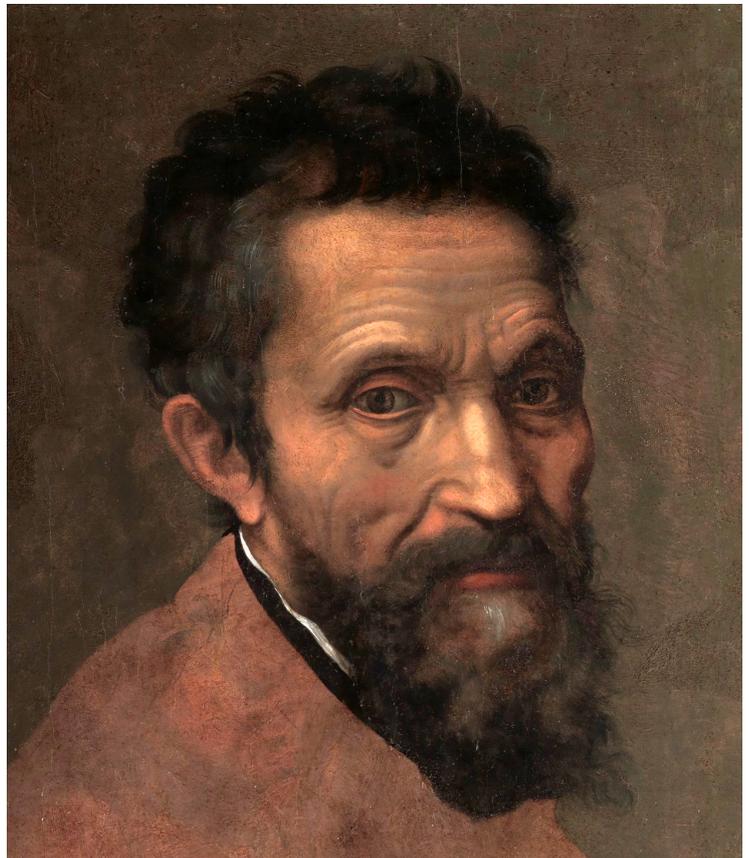

**Fig. 1. Michelangelo.** *Though perhaps best known as a sculptor and painter, Michelangelo di Lodovico Buonarroti Simoni was also an architect and a poet. Credit: Credit: Daniele da Volterra, Metropolitan Museum of Art, online collection (The Met object ID 436771)*

The importance of Lesson 1 cannot be overstated, and needs to be better appreciated by the scientific community. Scientists are driven by scientific research and our decadal surveys are science-focused. Yet within the world of politics, science is but one factor. If planetary scientists want to someday send a flagship mission to an Ice Giant, the essential requirements of Lessons 0 and 1 must be the mantras for getting it there. Successful mission advocates are always cognizant of the real landscape of funding and politics, and position missions accordingly.

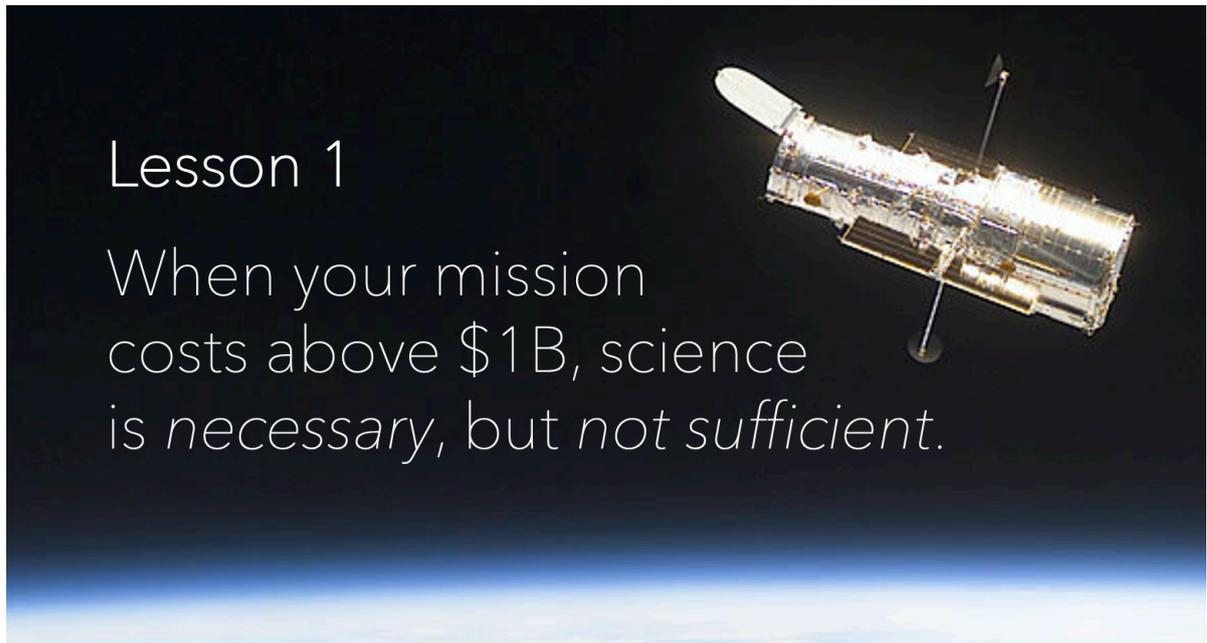

**Fig. 2. A crucial lesson for large missions.** *When your mission costs above $1 Billion, science is necessary but not sufficient. This amalgam comes from conversations with leaders of several large (multi-billion-dollar) science projects. This lesson was keenly exemplified by the Hubble Space Telescope, which—despite a forced linkage to NASA's human spaceflight and shuttle programs, as well as an ignominious start with a flawed mirror—is now recognized as the most scientifically-productive space mission ever launched.*

## Process

Planetary scientists often focus solely on the science return from a mission. Science is indeed essential and is the topic of many of the companion papers in this issue. However, Lesson 1 of Fig. 2 cautions us that factors external to science must be considered. I discuss here aspects of the "process" of missions and how they are executed, both in terms of the functional hardware and in terms of the management structures that give rise to the completed mission.

One evergreen lesson is to be open to new technologies. The London Ice Giants meeting featured speakers with a remarkable array of technological expertise (see companion papers in the issue). These technologies are important not only because of the science they enable, but because the companies and industries that support their development can be powerful allies in advocating for a future mission (see words in Fig. 2). Ice Giant missions are in a particular bind, though, because of the multi-decadal timescales required for completion. When only one mission is launched every thirty years, planners are unwilling to take risks on innovation. That said, technology has advanced so much in the thirty-plus years since Voyager that, for an Ice Giant mission, even older low-risk technology is far advanced beyond that of the Voyager era.

A second but less obvious lesson is to be open to innovative development processes championed by key individuals within Agencies or by industry. One experiment from the past illustrates this lesson: the Cassini Resource Exchange. This innovative process managed the development of the Cassini Spacecraft science payload (Fig. 3). A "market-based" economic system, it permitted the varied instrument teams to "trade resources" rather than being limited to a fixed allocation [1]. Initially, the tradable "commodities" in the Cassini Resource Exchange were data rate, budget, mass, and power. Later, cost was added to help with phasing issues. In the end, the program was able to build and deliver all of the proposed science instruments on time, the overall cost of the science payload grew by less than 1%, and as a bonus, the science payload mass actually shrank by 7%. Since a future flagship to an Ice Giant will likely have a complex and multi-faceted science payload, it might be worth revisiting a similar development process.

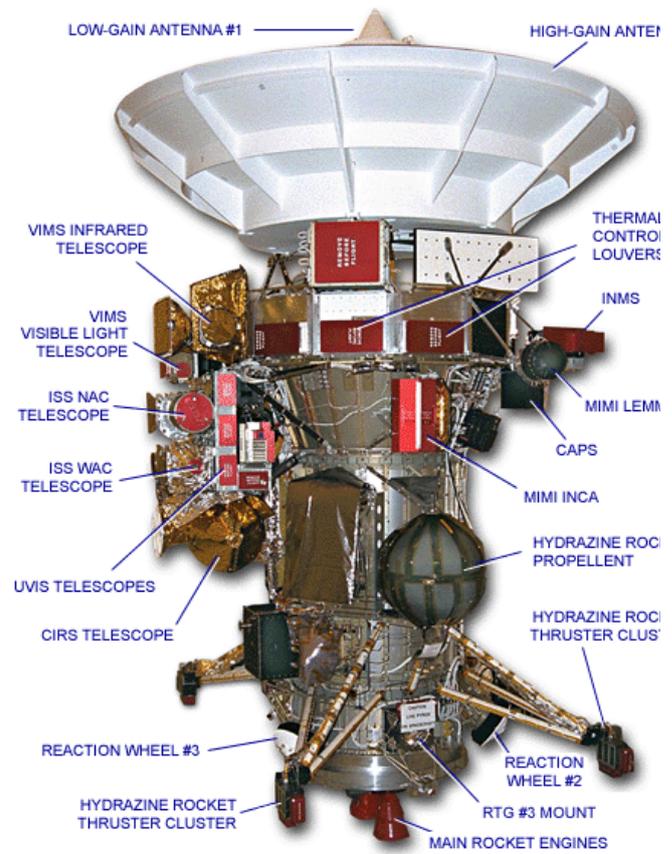

*Fig. 3. Cassini Mission to Saturn.* The Cassini program used an innovative "Resource Exchange" to facilitate the development of its science payload. Image source: https://solarsystem.nasa.gov/basics/cassini/.

Thirdly, planetary science does not operate in a vacuum. As of this writing, the US astrophysics community is deeply engaged in their decadal survey planning process. As a part of that activity, NASA commissioned four major flagship studies (collected at greatobservatories.org), and a significant effort has gone into assessing the process aspects of major (high-cost and high-complexity) missions. Table 1 (extracted from the LUVOIR mission presentation to the Astrophysics Decadal Survey [2]) illustrates a selection of papers that focus specifically on the issue of cost- and schedule-management for large NASA missions. The papers span astrophysics and planetary science as well as the US Department of Defense and the US Navy. The strategies and lessons in these documents have general applicability and are certainly relevant to Ice Giant missions.

When considering process for operating missions, consider alternative modes of operations. The New Horizons team implemented a "sleep" mode during the long cruise phases of the mission, when instruments were shut down and staff were reduced to skeleton crews. An alternative tactic for long cruise phases is to partner with outside groups to conduct ancillary science; e.g., conduct observations of transiting exoplanets, or make measurements of heliospheric conditions. Partnerships must be very carefully crafted, however, to avoid "scope creep" in the fundamental mission science in order to contain cost and maintain schedule.

Last but not least, it is highly advantageous for mission teams to engage early and often with industry counterparts. Avoid waiting until NASA has defined mission criteria or puts out a call; rather reach out to potential partners very early in the development process. While some may advocate for mission development to be focused within NASA Centers, there are significant advantages to having a broader base of support for a mission concept even in its earliest stages (one of the additional factors from Fig. 2).

Table 1

Research* on strategies for change in cost- and schedule-efficient project management

† Bitten, R., et al., 2019, Challenges and Potential Solutions to Develop and Fund NASA Flagship Missions, IEEE, 978-1-5386-6854-2/19

Wiseman, J., 2015, The Hubble Space Telescope at 25: Lessons Learned for Future Missions, IAUGA 2258532W

Mitchell, D., 2015, An Overview of NASA Project Management, MAVEN Magic, and Lessons Learned

Martin, P., 2012, NASA's Challenges to Meeting Cost, Schedule, and Performance Goals, OIG Report IG-12-021

† Feinberg, L., Arenberg, J., et al., 2018, Breaking the Cost Curve: Applying Lessons Learned from the JWST Development to Build More Cost Effective Large Space Telescopes in the Future, SPIE 10698-23.

Arenberg, J., Matthews, G., et al., 2014, Lessons We Learned Designing and Building the Chandra Telescope, SPIE 9144-25

CRS 2007, Defense Procurement 2004-2007: Full Funding Policy – Background, Issues, and Options for Congress, CRS Report for Congress, RL31404

O'Rourke, R., 2006, Navy Ship Procurement: Alternative Funding Approaches – Background and Options for Congress, CRS Report for Congress, RL32776

# Planning

One of the most challenging aspects of Ice Giant missions is multi-decadal timescales. This extended timeframe affects all aspects of mission planning and development, from science, to mission structure, to development cycles, and more (including people, but I defer that discussion to the next section). Here I touch on a few of these aspects, but this is not an exhaustive assessment.

Science, by definition, is the systematic study of the structure and behavior of the physical and natural world through observation and experiment. For our two local Ice Giants, we are scientifically stymied because the Voyager flybys in the 1980s remain our only brief close-up assessments of each system, and the instrumentation was limited in comparison with today's capabilities. Furthermore, these systems are so distant that the most powerful telescopes on Earth and in space are required to study their most basic aspects. Those telescopes are in high demand by other users, resulting in extremely limited time for Ice Giant research.

That said, nearly everything we know about these systems has changed since those brief Voyager encounters more than 30 years ago. The science results at the London Ice Giants Conference (see companion articles) are a testament to the perseverance and perspicacity of my colleagues, who have pushed to the very frontiers of astronomical instrumentation to glean this new knowledge.

## *Decadal Surveys*

The actual process of planning missions in the United States is rooted in the "decadal survey" process run by the US National Academy of Sciences. For two decades now, an Ice Giant mission has been highly ranked in our planetary decadal surveys in 2003 [11] and 2013 [12]. However, NASA's science budget is finite, and generally only one or at most two major (flagship) missions are initiated in each decade. To date, Ice Giants missions have not yet made the cut.[1] In order to move forward, it is imperative that an Ice Giant System mission be the top-ranked flagship class mission in the next planetary decadal.

Past decadal surveys, and past arguments about why to return to this or that planet, are often location-based: "we've done orbiters at planets X and Y, now it is time for Neptune" or "we've already done Z missions to that inner Solar System planet, let's go to Uranus instead." Our decadal surveys have encouraged that type of discourse through their very structure, with panels being focused on targets (Venus, or Mars, or Giant Planets). "Destination" thinking may no longer be scientifically sound anymore; as discussed in the next section, a broader scientific perspective is needed.

---

[1] As of this writing, NASA announced selections for its Discovery line. In an exciting development, the Trident mission was selected. Trident's science, however, is narrowly focused on Neptune's moon Triton, with little room to add science relevant to other aspects of the Neptune system without breaking cost and schedule caps, leaving out crucial data such on Neptune's atmosphere, interior, magnetic field, Neptune's complex and evolving ring system, and other moons.

*Diversity of Worlds*

Planetary science today has exploded with the confirmation of thousands of planets around other stars (**Fig. 4**). These "exoplanets" have not only been detected, but assessments are now beginning of their bulk properties, atmospheres, phase curves, and other measurements (see [13,14]).

It is fair to say that the theoretical models for understanding exoplanets currently outstrip the available data. A more robust scientific planning process will put our Solar System bodies solidly within the context of all planets. We could then ask very different types of questions. What can greater knowledge about the objects in our Solar System uniquely contribute to our understanding of planetary science writ large? What local measurements are crucial for refining theories for all planetary systems? What does the field of exoplanets tell us about planets in general, and what ground truths can missions to our local planets in turn provide?

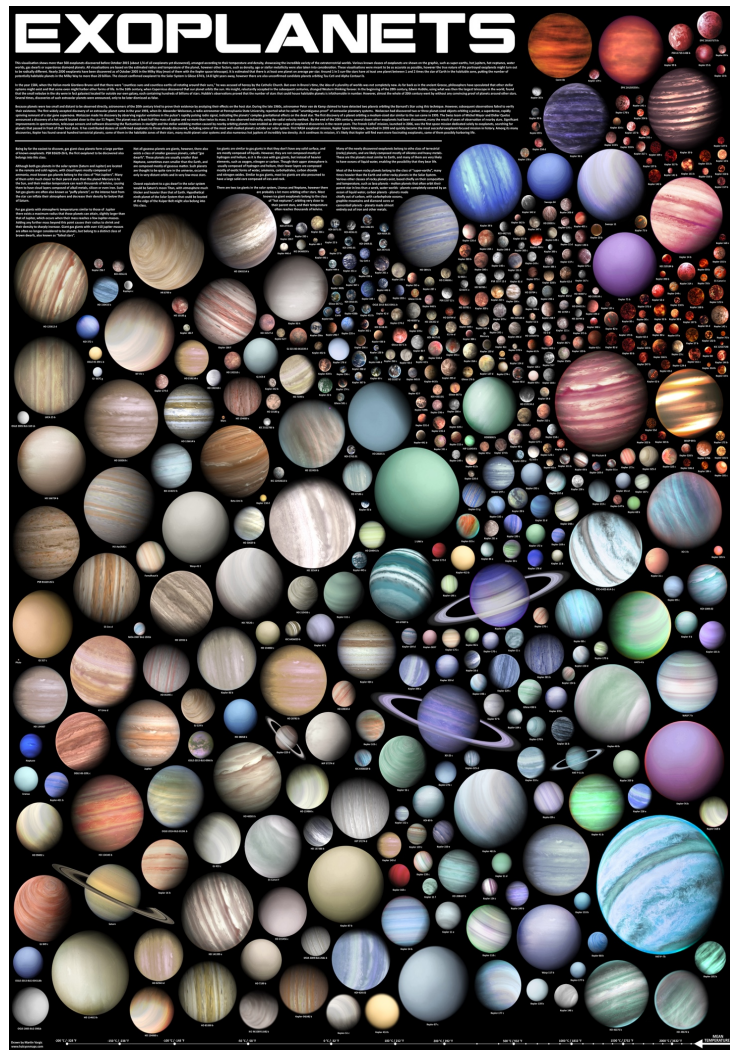

**Fig. 4. Planets galore.** *This poster illustrates the diversity of planets both within our Solar System and around other stars. Visualization by Martin Vargic.*

A number of talks at the London Ice Giants conference explicitly addressed the issue of Ice Giants in the context of exoplanets (e.g., [13]). If our next US planetary decadal is mindful of the broad scope of planetary science, and ensures that its recommendations are relevant and useful to those who study planetary systems elsewhere, then a comprehensive Ice Giant mission could very well rise to the top-ranked position. Its science yield per dollar would far outstrip many (if not all) other missions (but see Lesson 1 in Fig. 2). In the past decadal survey, a Uranus mission surprised many by rising to an extremely high rank on exactly this criterion.

*Actual Mission Funding Process*

Many scientists assume that if a mission is top-ranked in a decadal survey, then it is time to celebrate! In reality the process is far more multi-faceted. Within the United States, the federal funding process is complex. The power of the purse formally resides in the U.S. Congress and (almost more importantly for science) in key Congressional committees. The Executive branch of the US government (including the President's administration and funding agencies such as NASA) nevertheless exerts strong influences on funding priorities ("the President proposes, the Congress disposes"). In addition to this inherent "push and pull" tension in government activities, there are many external influences as well (see **Fig. 5**). It is not usual for a scientist to be asked by a Congressional staffer: "Why is your multi-billion-dollar space mission more important to my state than the federal facility that Agency X is proposing to build there?" Scientific priority and prowess, while necessary, are far from sufficient for success (see Fig. 2, again).

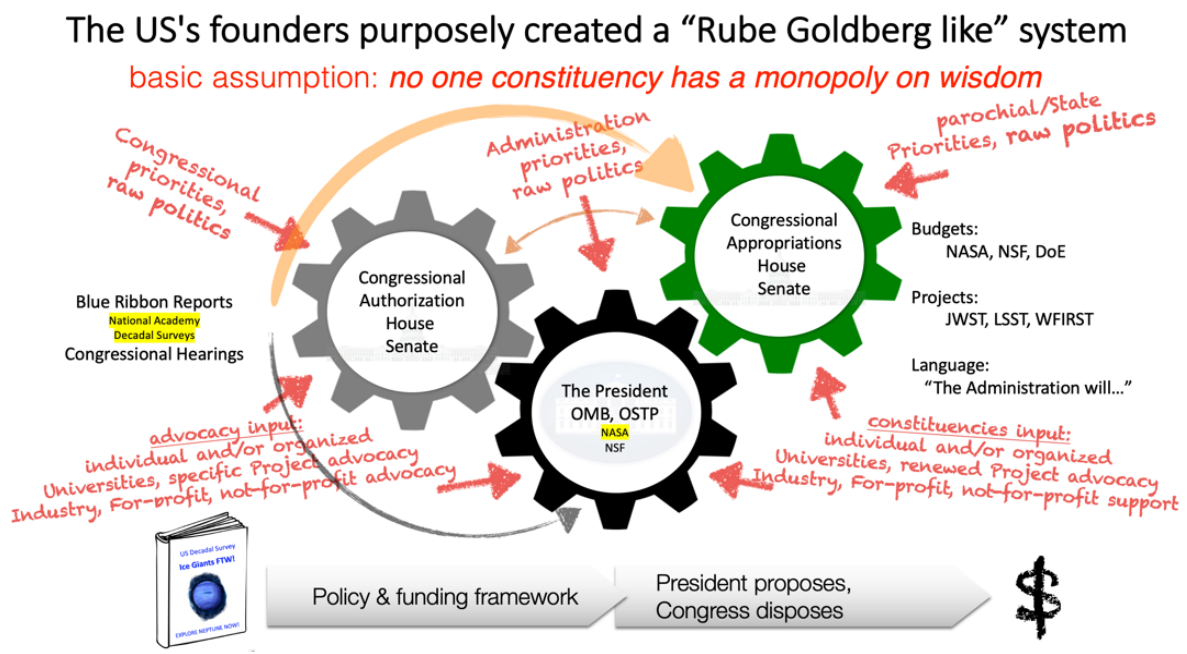

**Fig. 5. The US federal funding process.** *This complicated cartoon illustrates some of the many factors at play in determining US budgets. A top-ranked mission in the decadal survey (icon in lower left) does not instantly translate into funding for a NASA mission ($ in lower right), as many diverse constituents make their needs known. The small fonts of the US Decadal Survey and NASA (in yellow) allude to their relative importance in the grand scheme of US federal budgets. Modified from an image by Matt Mountain (AURA), and used with permission.*

The fundamental premise of Lesson 1 in Fig. 2 ("science is necessary but not sufficient") **likely applies internationally**, particularly for trans-national agencies such as ESA that work to balance the scientific and technological opportunities across their communities.. Conferences like the London Ice Giants Workshop are crucial for organizing the science community and aligning the science priorities. However, scientists from all countries should acknowledge the dual realities of Fig. 2's lesson and of a complicated political process analogous to that of Fig. 5.

At the London Workshop an excellent address from Dr. Fabio Favata about the European Space Agency (ESA) and its potential interest in Ice Giant exploration likened the relationship between ESA and NASA to a tango. Yet the realities of mission development strain this elegant analogy. For example, ESA runs its S, M, and L-class mission proposal calls, as well as "thematic calls" like Voyage2050. Similarly but asynchronously, NASA runs its Discovery, New Frontiers and Flagship missions, prioritized through the US decadal survey process. It's challenging to keep the disparate timelines aligned, meaning that one partner waits for the other to lead (or worse, both try to lead!). The lack of synchronicity has doomed many a proposal to failure because the "programmatics" haven't been adequately considered. Given this reality, coupled with the additional factors from Fig. 2 and the complexities inherent in governmental funding as per Fig. 5, a more appropriate dance analogy might be a "mosh pit" with many different dancers. To maneuver a mission with multiple participants to successful completion, scientists need to be aware of all the competing influences swaying their efforts. Greater (and transparent!) dialogue among agencies could mitigate the mosh, and facilitate more effective collaboration for international missions.

# People

Highly skilled and trained teams are a crucial component of planetary exploration. Behind every successful mission, there are anywhere from dozens to hundreds or even thousands of dedicated individuals, who together are jointly responsible for the design, build, test, and execution of a mission. For missions to the outer solar system, it usually takes decades to get from the initial planning to the final science. These timescales require a different kind of thinking by the leaders, a level of altruism and a recognition that they may not be building these missions for themselves, but rather for the next generation of scientists.

### *To Build a Cathedral*

I illustrate this change in thinking with the phrase "to build a cathedral" in two distinct ways. The first recounts the parable of the bricklayers, which in some tellings refers to the rebuilding of St Paul's Cathedral by Christopher Wren[2]:

> *One day in 1671, Christopher Wren observed three bricklayers on a scaffold, one crouched, one half-standing and one standing tall, working very hard and fast. To the first bricklayer,*

---

[2] *https://sacredstructures.org/mission/the-story-of-three-bricklayers-a-parable-about-the-power-of-purpose/*
Phil. Trans. R. Soc. A.

> *Christopher Wren asked the question, "What are you doing?" to which the bricklayer replied, "I'm a bricklayer. I'm working hard laying bricks to feed my family." The second bricklayer, responded, "I'm a builder. I'm building a wall." But the third brick layer, the most productive of the three and the future leader of the group, when asked the question, "What are you doing?" replied with a gleam in his eye, "I'm a cathedral builder. I'm building a great cathedral to The Almighty."*

Fully functional outer Solar System missions are cathedrals, not individual bricks or walls or even small chapels. It takes strong leadership to maintain that "gleam in the eye" of all those working on the project, but it is crucial for mission success.

Secondly, building outer solar system spacecraft is like cathedral building in its timescales: St. Paul's rebuilding took 35 years, but it was not an anomaly: Westminster Abbey's first major building lasted 25 years; Notre-Dame de Paris took about 100 years for its initial construction; and York Minster took a remarkable 252 years to complete. That is actually longer than a Neptune year (165 years)! **But just as cathedrals benefited the many generations that followed, our first dedicated mission to an Ice Giant will help shape planetary science (and the people within) for decades to come.**

## *Empower the youth*

Given the aforementioned timescales, we must recognize that we need to nurture the next generation of scientists. To put this in perspective, I note that I was in elementary school when Voyager was developed; and in high school when Voyager launched. It flew to the outer solar system while I was in college and graduate school. The Neptune flyby took place when I started post-doctoral research at the Jet Propulsion Lab. We do not yet know which children in school today will be the scientists on our Ice Giant mission in 25 years. But they are the ones for whom we build this mission!

Nurturing the next generation is challenging when there are no missions. Fortunately, significant progress in telescope observations over the past three decades has sustained the Ice Giant community. The remarkable capabilities of the Hubble, Spitzer, and Herschel Space Telescopes, the power of adaptive optics at Keck, Gemini, and VLT; radio and sub-millimeter data from the VLA and ALMA respectively — all these and more have propelled our Ice Giant system knowledge forward and given young scientists new data to explore. Similarly, advances in computation capability have allowed reanalysis of Voyager data and development of sophisticated new models. We also look toward the launch of James Webb Space Telescope, with its ability to study these systems at near- and mid-infrared wavelengths: Ice Giant observations are represented inWebb's Guaranteed Time Observations.[3]

---

[3] *For James Webb Space Telescope's Guaranteed Time Observations, Programs 1248 and 1249 focus on the atmospheres of Uranus and Neptune respectively, and Neptune's moon Triton will be studied under Program 1272. Descriptions of the observations can be found at: https://www.stsci.edu/jwst/phase2-public/1248.pdf,*

# Looking to the Future

The United States has initiated the process of preparing its next Decadal Survey in Planetary Science. This exercise gathers the community's current understanding of the state of the art in the field, and collectively crafts a program of exploration for the subsequent decades.  Ice Giant missions have regularly appeared in the past several planetary decadal reports but haven't risen to top billing. The changing nature of planetary science, and its expansive growth, coupled with significant advances in technology, could mean that this is finally the time that an Ice Giant mission could rise to a top-priority ranking.  The science yield would be tremendous, and our perspective on the Ice Giant systems would be irrevocably changed again, as it was more than 30 years ago by the Voyager 2 spacecraft.

One final lesson is to think from the perspective of the future.  When we plan for decades in advance, we have to place ourselves in that future world and imagine the perspective of the future scientists. Those future scientists will look back fondly at the wisdom and prescience of today's scientific community, who did not extrapolate from the (understandable) pessimism of the moment, and instead envisioned a different vibrant future of outer Solar System exploration. The scientists of the 2020s heeded Lesson 0 from Michelangelo; acted within the precepts of Lesson 1 to build a broad and multi-faceted consortium of supporters; developed innovative processes; planned wisely; and prepared the next generation of scientists for Ice Giant exploration. As a result, our future scientists are again exploring the skies of Neptune, the cryovolcanoes of Triton, the weird rings of Uranus, and the myriad moons of these distant worlds (**Fig. 6**).

---

*https://www.stsci.edu/jwst/phase2-public/1249.pdf, and https://www.stsci.edu/jwst/phase2-public/1272.pdf.*

*Phil. Trans. R. Soc. A.*

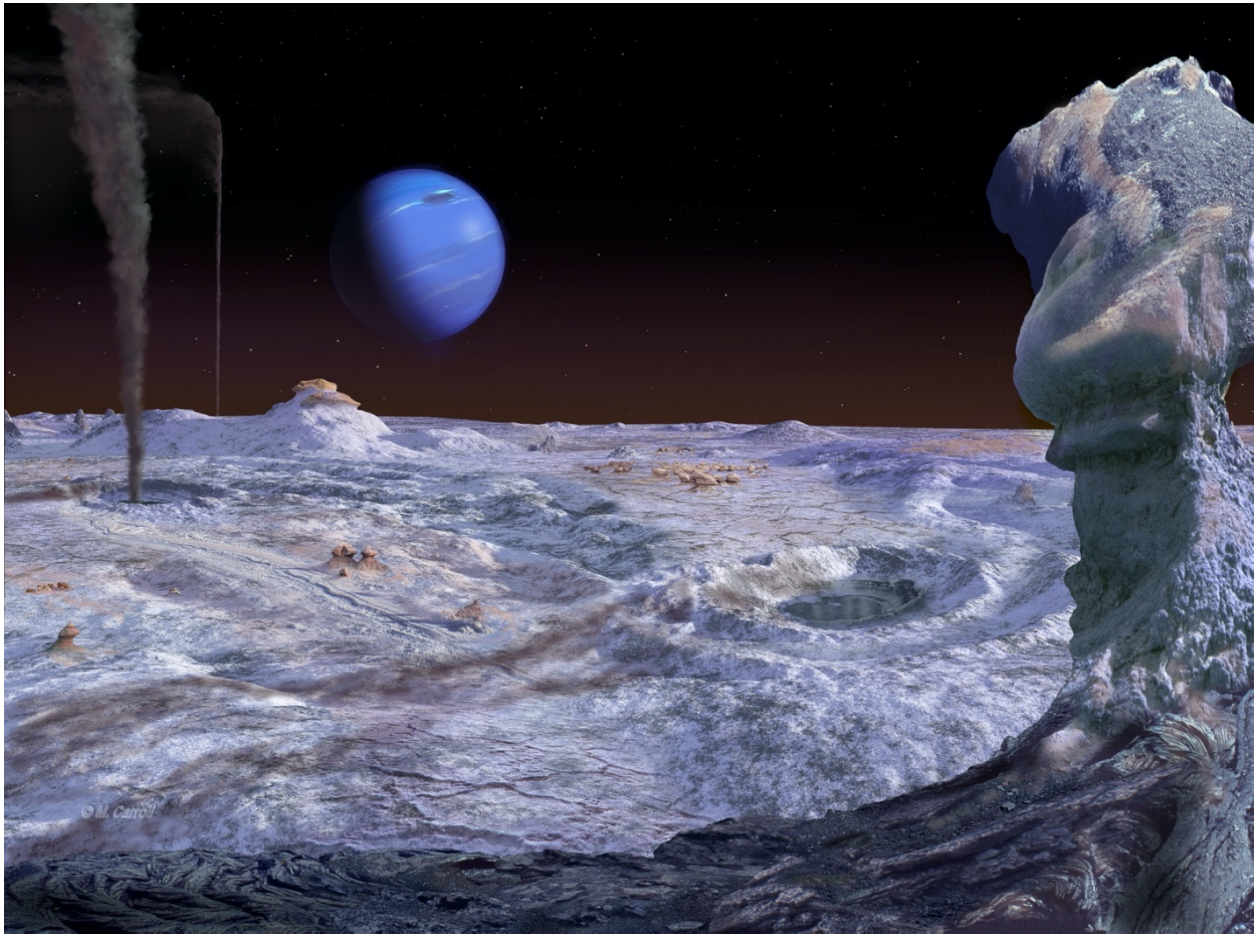

*Figure 6. Artist's concept of Neptune as seen from its large moon Triton. Used with permission from the artist, Michael Carroll ([www.stock-space-images.com](www.stock-space-images.com)).*

# Additional Information

**Data Accessibility.**  This article contains no data sets.

**Author Contributions.**  H. B. Hammel (sole author) drafted the manuscript.

**Competing Interests.**  The author has no competing interests.

**Funding Statement.** The author is funded solely through her home institution.

**Acknowledgments.**  The author sincerely thanks Dr. Leigh Fletcher and the Science Organizing Committee of the London Ice Giants Workshop for the opportunity to present these lessons. Thanks also go to the Royal Society for their exquisite hosting of the Workshop.  Finally, the author thanks all her colleagues from the Voyager mission, especially Candice Hansen-Koharcheck, Richard Terrile, David Grinspoon, and John Casani.

*Phil. Trans. R. Soc. A.*